\documentstyle [preprint,aps]{revtex}
\newcommand{\be}{\begin{equation}}
\newcommand{\ee}{\end{equation}}

\begin {document}
\draft

\title{Transverse force on a quantized vortex in a superfluid}

\author {D.J. Thouless$^1$, Ping Ao$^{2,1}$ and Qian Niu$^3$}

\address{$^1$Department of Physics, Box 351560,
    University of Washington, Seattle, WA 98195,\\
$^2$Department of Theoretical Physics, 
   Ume\aa \ University, Ume\aa, Sweden, and\\
$^3$Department of Physics, University of Texas, Austin, TX 78712  }
\date{\today}
\maketitle

\begin {abstract}
 { We have derived an exact expression for the total 
nondissipative transverse force acting on a
quantized vortex moving in a uniform background. 
The derivation is valid for neutral boson or fermion superfluids,
provided the order parameter is a complex scalar
quantity. This force is determined by the one-particle
density matrix far away from the vortex core, and is found to be the 
Magnus force proportional to the superfluid density.  We conclude that 
contributions of the localized core states do not change this force.  }
\end {abstract}

{\bf Phys. Rev. Lett.} (in press)

\narrowtext

For a vortex moving in a superfluid there is a force transverse to
its velocity which is the counterpart of the Magnus force in classical
fluid mechanics \cite{lamb}.  This was used in the experiment by which
Vinen \cite{vinen} established the quantization of circulation round a wire
in $^4$He.  The obvious generalization of the form of the Magnus force to
the situation in superfluid helium is to take the force per unit length to
${\bf F}_M=\rho_s {\bf K}\times({\bf v}_V-{\bf v}_s)$, where ${\bf K}$ is
the circulation, ${\bf v}_V$ is the vortex velocity, and $\rho_s$ is the
superfluid density and ${\bf v}_s$ is the superfluid velocity in the
neighborhood of the vortex.  Even for the case of $^4$He there has been
some controversy about the form of this transverse force, despite the
experimental measurements of Vinen \cite{vinen} and
of Whitmore and Zimmermann \cite{zimmer} at temperatures where there is an
appreciable difference between $\rho$ and $\rho_s$. This controversy is
partly based on the observation by Iordanskii \cite{iordanskii} that
excitations such as phonons are asymmetrically scattered by a vortex, and
this should lead to a transverse force proportional to $\rho_n$ and to the
velocity ${\bf v}_n-{\bf v}_V$ of the normal fluid component relative to
the vortex. The magnitude of this term is not immediately obvious, nor is
it obvious whether this contribution should be added to the Magnus force
as we have written it here, or to the expression for ${\bf F}_M$ with the
superfluid density replaced by the total fluid density. It has been 
convincingly argued \cite{donnelly} that the correct form of the 
transverse force is
\be
{\bf F} =\rho {\bf K}\times({\bf v}_V-{\bf v}_s)+\rho_n {\bf K}\times({\bf
v}_n-{\bf v}_V)  \;,
\label{eq:donnelly}\ee
which can also be written as ${\bf F}_M$ plus a term proportional to ${\bf
v}_n-{\bf v}_s$.  The problem of determining the correct form of the
force is discussed in detail by Sonin \cite{sonin}.

For a fermion superfluid there appear to be additional complications
introduced by the spectrum of low-lying fermion excitations associated with
the vortex core.  Volovik \cite{volovik93,volovik95} has identified an
additional contribution proportional to ${\bf v}_n-{\bf v}_V$ associated
with these excitations, which, in some estimates, cancels most of the
Magnus force.  However, measurements of the circulation
quantization in the B phase of superfluid $^3$He \cite{packard} show a Magnus
force of the expected size.  In superconductors there seems to be little
unambiguous evidence about the existence or magnitude of this transverse
force. 

In earlier work we have shown the close connection between the Magnus
force and a Berry phase \cite{ANT}, but our arguments were dependent on a
detailed model of the vortex structure.In this paper we consider a
single, isolated vortex forced to move through an infinite superfluid with
uniform velocity ${\bf v}_V$ by some moving pinning potential.  We
are able to make an exact evaluation of the coefficient of ${\bf v}_V$ in
the force which has to be applied to the vortex in terms of the integral
of the momentum density round a large loop that surrounds the vortex,
and this is essentially the term proportional to ${\bf v}_V$ in
eq.\ (\ref{eq:donnelly}).  The first step in the argument involves a
perturbative calculation of the effect of the motion of the vortex in
terms of the instantaneous eigenstates of the Hamiltonian.  In the case of
a system with no inhomogeneous substrate this expression can be written
in terms of a Berry phase \cite{geometric}.  This expression can be
written in terms of the derivatives of one- and two-particle Dirac density
matrices, or, alternatively, as the commutator of two components of the
total momentum operator.  This can be turned into an integral over a
surface (or a loop for the two-dimensional case) surrounding the vortex
and at a large distance from it.  From this we can conclude that there
are no corrections to the term proportional to ${\bf v}_V$ in
eq.\ (\ref{eq:donnelly}). 
   
For simplicity we work with a two-dimensional system, but the argument can
readily be applied to a three-dimensional system. We assume that the system
is homogeneous and infinite. The isolated vortex in the superfluid can be
pinned to a position ${\bf r}_0$ by applying a potential $\sum_iV({\bf
r}_0-{\bf r}_i)$ which acts on all the particles in the superfluid.
For a static problem this potential can be arbitrarily weak, but when
the pinning potential is made to move through the superfluid it must
have a strength sufficient that the vortex does not get detached from
the pinning center, either by tunneling with the aid of the Magnus
force, or by acquiring enough energy from the phonon system.  The
potential has to be strong enough to to break the degeneracy of this
broken symmetry vortex state and allow us to use a perturbative
treatment of the velocity.  The situation is somewhat reminiscent of
the theory of the Stark effect in atoms, where a perturbative
treatment of a weak uniform electric field is possible, despite the
fact that the electric field connects the bound levels of the
electrons in the atom to a continuum of ionized states of the same energy.
The pinning potential may also be very strong, as it is in the Vinen 
experiment, where the circulation is pinned to a solid wire.  Our
results are quite independent of the details of the pinning
potential, provided that it is strong enough to allow perturbative
methods to be used, and does not break cylindrical symmetry.

In terms of the instantaneous eigenvalues $E_\alpha(t)$ and eigenstates
$|\Psi_\alpha(t)\rangle$, for which we choose phases such that
$\langle\Psi_\alpha|\dot\Psi_\alpha \rangle=0$, the time-dependent
solutions of the Schr\"odinger equation can be written as \be
   |\Psi(t)\rangle= a_\alpha(t)e^{-i\int^t E_\alpha(t')dt'/\hbar}
   |\Psi_\alpha(t) \rangle +\sum_{\beta\ne\alpha}
     a_\beta(t)e^{-i\int^t E_\beta(t')dt'/\hbar} |\Psi_\beta(t)\rangle \;, 
\ee
where, to first order in the velocity, $a_\alpha(t)=1$, and
\[
  a_\beta(t)=-\int^tdt' \langle\Psi_\beta|\dot\Psi_\alpha \rangle
   e^{-i\int^{t'}(E_\alpha- E_\beta)dt''/\hbar}
\]
\be
   =-\int^tdt' \langle\Psi_\beta|{\bf v}_V\cdot \nabla_0 \Psi_\alpha
    \rangle e^{-i\int^{t'}(E_\alpha- E_\beta)dt''/\hbar}\;.
\ee
Here $\nabla_0$ denotes the partial derivative with respect to
the position ${\bf r}_0$ of the pinning potential. This gives the
expectation value of the force on the pinning potential as
\[
  {\bf F}= -\sum_\alpha f_\alpha \langle\Psi_\alpha|
   \nabla_0H|\Psi_\alpha \rangle
\]
\be
   +\sum_\alpha f_\alpha \langle\Psi_\alpha|
    \nabla_0H  e^{i\int^{t}(E_\alpha- H)dt'/\hbar}
   {\cal P}_\alpha \int^tdt' 
   e^{-i\int^{t'}(E_\alpha- H)dt''/\hbar}{\bf v}_V\cdot \nabla_0 +{\rm
   h.c.} |\Psi_\alpha \rangle \;,
\label{eq:kubo}\ee
where $H$ is the total Hamiltonian of the system, which depends on
${\bf r}_0$ only through the pinning potential, $f_\alpha$ is the
occupation probability of the state $\alpha$, ${\cal P}_\alpha$ is
the projection operator off the state $\alpha$, and h.c. denotes the
Hermitian conjugate term. The first term on the right of this equation
includes any forces on the vortex proportional to the normal and
superfluid velocities, and we have not attempted to evaluate
these. We concentrate on the second term, which gives those forces 
proportional to the vortex velocity. There are energy-conserving
dissipative forces included in this expression, but, as with the Kubo
formula for longitudinal conductivity, their evaluation requires a
careful limiting process, and we have not done this here. We
concentrate on the contributions from intermediate states whose energy
is different from that of the initial state, and it is these that give
rise to the transverse force perpendicular to the vortex velocity. 

For the homogeneous system that we are considering the eigenvalues are
independent of time, so eq.\ (\ref{eq:kubo}) can be written in the
simpler form
\be
   {\bf F}= -\sum_\alpha f_\alpha \langle\Psi_\alpha| \nabla_0
                      H|\Psi_\alpha \rangle
            +\sum_\alpha f_\alpha 
            \left\langle \Psi_\alpha \left|
      \nabla_0H {i\hbar{\cal P}_\alpha\over E_\alpha- H} {\bf v}_V\cdot
        \nabla_0 +{\rm h.c.} \right| \Psi_\alpha \right\rangle \;.
\ee
Since $\nabla_0H$ is the commutator of the operator $\nabla_0$
with $H$, the commutator cancels the energy denominator, and the part 
linear in the vortex velocity can be written as
\be
   {\bf F}\times \hat{\bf n}=-i\hbar{\bf v}_V\sum_\alpha f_\alpha
     \left(\left\langle {\partial\Psi_\alpha\over \partial x_0} \left|
    {\partial\Psi_\alpha\over \partial y_0}\right\rangle -\left\langle
    {\partial\Psi_\alpha\over \partial y_0} \right| {\partial\Psi_\alpha\over
    \partial x_0}\right\rangle  \right) \;,
\label{eq:berryphase}\ee
where $\hat{\bf n}$ is the unit vector normal to the plane.
The Berry phase associated
with a closed loop in ${\bf r}_0$ can be written as the integral
over the area enclosed by the loop of $ F_M/\hbar v_V$.
Equation (\ref{eq:berryphase}) corresponds to the familiar form
for the Berry phase \cite{geometric}. 
Since we can choose the wave functions in such a way that the
dependence on ${\bf r}_0$ is entirely through $({\bf r}-{\bf r}_0)$,
the partial derivatives with respect to ${\bf r}_0$ can be replaced
by a sum over partial derivatives with respect to the particle
coordinates ${\bf r}_j$.  Upon thermal average, this expression can now be 
written in terms
of the Dirac density matrices as
\[
  F /v_V=-i\hbar\sum_\alpha f_\alpha \sum_{i,j}
   \left( \left\langle {\partial\Psi_\alpha\over \partial x_i} \left|
    {\partial\Psi_\alpha\over \partial y_j}\right\rangle - \left\langle
    {\partial\Psi_\alpha\over \partial y_j} \right| {\partial\Psi_\alpha\over
      \partial x_i}\right\rangle \right)
\]
\[
   =-i\hbar \hat{\bf n}\cdot \int\int d^2r[\nabla\times\nabla'\rho({\bf
     r}', {\bf r})]_{r=r'}
\]
\be
  -i\hbar \hat{\bf n}\cdot\int\int d^2r_1\int\int d^2r_2 [2 \nabla_1
  \times \nabla'_2 \Gamma({\bf r}'_1,{\bf r}'_2; {\bf
  r}_1,{\bf r}_2)]_{r=r'} \;,  
\label{eq:magnusforce1}\ee
where $\rho$ and $\Gamma$ are the one- and two-particle Dirac
density matrices for the system, and the sum over $i$ and $j$ denotes
a sum over all the particles in the system. 

The integral over the two-particle density matrix $\Gamma$ vanishes.  
Because of the symmetry between ${\bf r}_1$ and ${\bf r}_2$,
the integrand in this integral can also be written as 
\be
   [(\nabla_1+\nabla'_1)\times(\nabla_2-\nabla'_2) 
   \Gamma({\bf r}'_1,{\bf r}'_2; {\bf r}_1,{\bf r}_2) -\rho({\bf
r}'_1,{\bf r}_1) \rho({\bf r}'_2,{\bf r}_2) ]_{r=r'} \;;
\ee
the replacement of $\Gamma$ by $\Gamma-\rho\rho$ is also obvious,
since the integral over ${\bf r}_1$ of ${\bf grad}_1 \rho ({\bf r}_1)$
vanishes.  For any value of ${\bf r}_2$ this can be integrated over
${\bf r}_1$ to give a correlation function between distant points
which should vanish if the range of integration is
sufficiently large.

A more formal way of getting the same result is to say that the first
line of eq.\ (\ref{eq:magnusforce1}) gives the transverse force in terms
of the expectation value of the commutator of the $x$ and $y$
components of the total momentum of the particles.  This commutator is
a one-particle operator given by
\be
   [P_x,P_y]= \int\int dx\,dy\left({\partial \psi^{\dag}\over \partial x}
   {\partial\psi\over \partial y} -{\partial \psi^{\dag}\over \partial y}
   {\partial\psi\over \partial x} \right) \;,
\ee
where the $\psi^{\dag},\psi$ are creation and annihilation operators
for fermions or bosons. Since it is a one-particle operator, its
expectation value is given by the one-particle Dirac density matrix.

The integrand of the first term in eq.\ (\ref{eq:magnusforce1}) 
can be written as half the curl of $(\nabla'
-\nabla)\rho({\bf r}',{\bf r})$ evaluated at ${\bf r}={\bf
r}'$. Now divide the integrals over ${\bf r}$ and ${\bf r}_1$ into a sum
over finite areas labeled with an index $\sigma$; the first of
these areas will be centered on the position of the vortex, and the
others are all well away from the vortex. Stokes' theorem can be used
to write the result as 
\be 
   F / v_V= \sum_\sigma\oint_\sigma {i\hbar\over 2}d{\bf r}
    \cdot[(\nabla- \nabla')\rho({\bf r}', {\bf r})]_{r=r'}  \;. 
\label{eq:magnusforce2}
\ee
This result is exact, and there are no contributions to the
integrals from the neighborhood of the vortex core, since we have
chosen the boundaries of the regions of integration to be well away
from the core; any contributions from the core states or from the
properties of the core must be reflected in the density matrices
well away from the core. 
In particular, because localized states inside the 
vortex core, such as occur for an s-wave superconductor
\cite{caroli}, do not influence the one-particle density matrix far away
from the core, their contributions to the transverse force must be zero.
It should be pointed out that this property is not transparent in
eq.\ (\ref{eq:magnusforce1}), where both detailed forms of density matrices
and explicit integrations  are needed to recover it. 

For a neutral superfluid the integrand in eq.\ (\ref{eq:magnusforce2})
is just the momentum density, equal, by definition, to $\rho_s{\bf
u}_s+\rho_n{\bf u}_n$, where ${\bf u}_s$ and ${\bf u}_n$ are the local
values of the superfluid and normal velocities. If the circulation of
the normal fluid is zero, this term gives ${\bf F} = \rho_s {\bf
K}\times{\bf v}_V$, the Magnus force ${\bf F}_M$
\cite{vinen,packard,ANT}. From the Galilean invariance of the problem
we can deduce that ${\bf v}_V$ should be replaced by ${\bf v}_V-{\bf
v}_s$, but we cannot determine the coefficient of the term
proportional to ${\bf v}_n-{\bf v}_s$ without explicitly putting in
different background velocities for the normal fluid and the
superfluid components.  

At first sight it is surprising that we can obtain an exact result for
this problem, since there are very few problems to which quantum
many-body theory gives exact answers, but there is a similar argument that
can be used in classical hydrodynamics. If fluid is streaming past a fixed
vortex with velocity ${\bf v}$, one can calculate the momentum balance on
a very large cylinder centered on the vortex. The net pressure force per
unit length on the interior of this cylinder is $\rho_n {\bf K}\times{\bf
v}/2$, the momentum flow out from the cylinder is $-\rho_n {\bf
K}\times{\bf v}/2$, and these must be balanced by a force $-\rho_n {\bf
K}\times{\bf v}$ acting on the vortex core.  It is by an elaboration of
this argument that Barenghi {\it et al.} \cite{donnelly} obtain their
expression for the contribution of the flow of the normal fluid past the
vortex.

Our conclusion is that for any infinite, homogeneous, neutral
superfluid with a scalar order parameter the transverse force has the
form $\rho_s {\bf K}\times{\bf v}_V$, and those mechanisms which have
been suggested as giving corrections must actually give alternative
ways of looking at the same thing.  Because our argument is, like the
classical argument quoted in the previous paragraph, basically an
argument that balances the forces on the vortex core with forces and
momentum flow at large distances for the vortex, there is no obvious
generalization that allows for the effect of the disordered background
potential which is normally important for real superconductors.

 {\bf Acknowledgements:} We wish to acknowledge the hospitality of the
Aspen Center for Physics, where this work was conceived and
completed. We wish to thank Daniel Fisher, Carlos Wexler, Boris
Spivak, Xiao-Mei Zhu and numerous other people for helpful comments on
this problem. The work was supported by US NSF Grant No. DMR-9220733
and Swedish Natural Science Research Council(P.A.).  This paper is
respectfully dedicated to Joe Vinen, whose work first showed that the
transverse force on a vortex in superfluid $^4$He had this form.

\end{document}